# Comparative studies of hydrogen dissolution and release behavior in zirconate oxides by TDS and TMAP4 analysis

**M. Khalid Hossain[1, 2, *], K. Hashizume[1]**

[1]*Department of Advanced Energy Engineering Science, Interdisciplinary Graduate School of Engineering Science, Kyushu University, Fukuoka 816-8580, Japan.*

[2]*Atomic Energy Research Establishment, Bangladesh Atomic Energy Commission, Dhaka 1349, Bangladesh*

*Correspondence: khalid.baec@gmail.com, khalid@kyudai.jp

## Abstract

Proton-conducting oxides are potential materials for electrochemical devices such as fuel cells, hydrogen pumps, hydrogen sensors, and the tritium purification and recovery system in nuclear fusion reactors. The hydrogen concentration in oxide materials is important, but its precise measurement is difficult. In this study, thermal desorption spectroscopy (TDS) was used to investigate hydrogen dissolution and release behavior in proton-conducting oxides, yttrium (Y), and cobalt (Co) doped barium-zirconates in the temperature range of 673-1273 K using deuterium ($D_2$) and heavy water ($D_2O$). Specimens were prepared with conventional powder metallurgy: the powder of three zirconates, $BaZr_{0.9}Y_{0.1}O_{3-\alpha}$ (BZY), $BaZr_{0.955}Y_{0.03}Co_{0.015}O_{3-\alpha}$ (BZYC), and $CaZr_{0.9}In_{0.1}O_{2.95}$ (CZI) was pressed into discs and fired in the air at 1873 K for 20 h. The densities of the sintered BZY, BZYC, and CZI specimens were 98%, 99.7%, and 99.5% of the theoretical densities respectively. XRD, SEM, and EDX were performed for the structural and morphological analysis of the sintered samples. From TDS measurement, a similar trend of temperature-dependent hydrogen solubility was obtained for all samples compared to the tritium imaging plate (TIP) method's literature data of HT- and DTO-exposed samples. To compare the experimental results of the deuterium desorption profile derived by TDS analysis, the simulation code of the tritium migration analysis program, version 4 (TMAP4) was employed.

## 1 Introduction

Hydrogen has been a promising clean energy resource for mitigating global climate issues and the demand is increasing gradually to ensure energy security and less air pollution. Hydrogen is going to be a good alternative to traditional fossil fuels due to the development of fuel cells for mitigating global climate change [1]. The major production of hydrogen (~95%) comes from the separation and purification of fossil fuels [2]. The low-cost membrane separation process enables almost zero-emission plants as this technique could produce hydrogen more effectively at a large scale with negligible impurities [3]. For this purpose, the separation of hydrogen from gases becomes promising as low-cost energy with the application of membrane technology based on Ni and Zr doped $BaCeO_3$ with stable in chemical reactions as the gas mixtures composed of $H_2O$, $CO_2$, and $H_2S$. In addition to nickel and barium cerate doped high proton conductors excel to develop cost-effective materials for hydrogen separating membranes [4,5].

Perovskite proton-conducting oxides ceramic membrane has drawn considerable attention for hydrogen separation applications [6] over the last few decades due to several advantages like thermal stability, good mechanical strength, high hydrogen permeation flux, and chemical resistivity in corrosive environments [7,8]. Owing to their proton-conducting properties at high temperatures (500 °C to 1000 °C) and high irradiation, perovskite-based ceramics have emerged as being useful in fusion reactor fuel purification systems, hydrogen pump flow systems, as well as fuel cells for energy generation [9]. The use of proton-conducting ceramics (PCC) including $SrCe_{0.95}Yb_{0.05}O_{3-\alpha}$ in tritium pumps has drawn the attention of nuclear fusion technology researchers because of its excellent performance in fusion reactors to recover the tritium molecules from water vapor of low concentration tritium in blanket sweep gas [10]. In a hydrogen pump selective extraction of the hydrogen isotopes is made possible by the application of optimum electrical potential difference which facilitates enrichment of the bred tritium gas from low partial pressure to high partial pressure side from the blanket sweep gas [10].

The requirements of being a good protonic conductor include- high proton conductivity with reliability and should have practical durability against steam or $CO_2$. As protons are not intrinsic to oxide lattices, they can be formed through positively charged hydroxide defects which requires a considerable concentration of oxygen vacancies in the oxides. Such observation is the key reason for zirconate-based perovskites' being high-performance proton conductors since the zirconium has a unique ability to create oxygen vacancies when doping with a trivalent cation ($Y^{3+}$, $In^{3+}$, $Yb^{3+}$, $Gd^{3+}$, etc.) [11–13]. Although barium zirconate-based perovskites showed high promises for their greater chemical stability and higher mechanical strength than the cerates, poor sinterability appeared as a major drawback which results in the high grain boundary densities and lowers the overall proton transport properties in barium zirconates, which could be solved by solid-state reactive sintering [14].

Therefore it is crucial to understand hydrogen isotope behaviors, especially the solubility and diffusivity of hydrogen in oxide materials, to use them as functional materials in fusion reactors [15–18]. However, hydrogen isotope behavior data in oxide materials are limited and scattered. Therefore, the motivation of this work is to carry out an in-depth study of hydrogen isotopes behaviors in oxide materials that could be suitable for the coating of metal piping in a fusion reactor's DT fuel circulation system to prevent tritium leakage [19,20] and also to study oxide materials in which hydrogen isotopes selectively pass through and can purify unused tritium during fusion reactions [21,22].

Tritium concentration (i.e., dissolution or solubility of hydrogen) has been estimated to be approximately 1.8 mol% in our prior works on barium cerate protonic conductor using tritium imaging plate technique [17]. In 10% Y-doped $BaCeO_3$ ($BaCe_{0.9}Y_{0.1}O_{3-\alpha}$) the concentration of tritium was nearly constant at a 400–600 °C temperature range. Since deuterium used in a nuclear fusion reactor together with tritium, and $BaCe_{0.9}Y_{0.1}O_{3-\alpha}$ shows improved proton conductivity [23–28], the deuterium dissolution, release, solubility, and diffusivity behaviors are evaluated in the deuterium gaseous atmosphere (e.g. $D_2$) and deuterium vapor (e.g. $D_2O$) in $BaCe_{0.9}Y_{0.1}O_{3-\alpha}$, as well.

Therefore, to achieve the above-mentioned objectives, the Y- and Co-doped barium zirconates (i.e., BZY and BZYC) and In-doped calcium zirconate (i.e., CZI) proton conductors were selected as the zirconate specimens in this study. Then, the following initiatives have been taken for these specimens, and the results with discussions are reported in this paper:

I. Two different deuterium sources ($D_2$ gas and $D_2O$ vapor) were exposed to the specimens for TDS experiments.

II. Hydrogen dissolution and release behavior, along with hydrogen solubility, were analyzed using the TDS method.

III. The hydrogen solubility data of the zirconates were obtained using the thermal desorption spectroscopy (TDS) method [29,30] and compared with our previously reported TIP solubility data **c**.

IV. Using the simulation code of the TMAP4, the experimental results of tritium diffusivity from the TIP method and deuterium desorption profile derived from the TDS method were compared [31,32].

V. The behavior of hydrogen isotopes in barium and calcium zirconates is compared in detail.

## 2 Experimental

### *2.1 Sintered samples preparation and characterization*

The raw zirconate materials were calcined powders of $BaZr_{0.9}Y_{0.1}O_{2.95}$ (BZY), $BaZr_{0.955}Y_{0.03}Co_{0.015}O_{2.97}$ (BZYC), and $CaZr_{0.9}In_{0.1}O_{2.95}$ (CZI) collected from TYK Corporation in Japan, to prepare sintered pellets of BZY, BZYC, and CZI. The details of sample preparation and polishing procedure are discussed in our previous reported papers, where samples were sintered at 1650 °C for 20 h in an air atmosphere [12,15,23,33–36]. The sintered-polished samples had a diameter of 7.5 mm, a thickness of 2.0–2.3 mm, and a weight of 0.6 g, respectively.

The sintered materials' bulk density and relative density (kg/m$^3$) were calculated using Archimedes' technique [15]. A powder X-ray diffractometer (MultiFlex, Rigaku Company Ltd.) was used to structurally examine all powder and sintered zirconate samples. A scanning electron microscope (SEM) (JSM-6010, JEOL, Japan) was used to examine the samples' morphology.

### *2.2 $D_2$ and $D_2O$ exposure*

The sintered disc zirconate specimens (BZY, BZYC, and CZI) were cut using a diamond saw into parallelepiped pieces (~7.7 × 2.2 × 0.5 mm$^3$) for $D_2$ or $D_2O$ exposure. For the TDS experiment, we used the following two sets (Sets 1–2) of samples for $D_2$ or $D_2O$ exposure:

**Set 1:** BZY, BZYC, and CZI samples exposed to $D_2$ gas (or $D_2$).

**Set 2:** BZY, BZYC, and CZI samples exposed to heavy water (or $D_2O$).

A schematic of the $D_2$- or $D_2O$-exposure apparatus is shown in **Figure 1**. For Set 1, the polished sintered ~0.5 mm thick BZY, BZYC, and CZI cut samples were exposed to $D_2$ gas (~1.33 kPa) at 873 K for 1 h to facilitate deuterium absorption into the samples. The exposure conditions of all samples of Set 1 are given in **Table 1**. The polished samples were put in a quartz tube, and the inside of the apparatus was evacuated to about $10^{-7}$–$10^{-8}$ Torr using a rotary pump and a turbomolecular pump to begin $D_2$ exposure. Before being exposed to $D_2$, the samples were vacuum annealed at 1273 K for 1 h to remove impurities. Following that, the samples were kept at a specific temperature range (673–1073 K). $D_2$ gas was injected at a pressure of 1.33 kPa into the sealed vacuum apparatus. After a predetermined exposure duration (1–4 h), deuterium was dissolved in the sample. After the $D_2$ exposure, the quartz tube containing the samples was quenched with room-temperature water to allow deuterium to dissolve in the samples. The $D_2$ gas that remained in the line was recovered, and the sample was subsequently removed. The dissolution and release of hydrogen were then analyzed using the TDS study. The exposure of $D_2O$ in the samples was the same as that of $D_2$ exposure in Set 1, but the exposure conditions were different. The exposure conditions of all samples of Set 2 are given in **Table 2**.

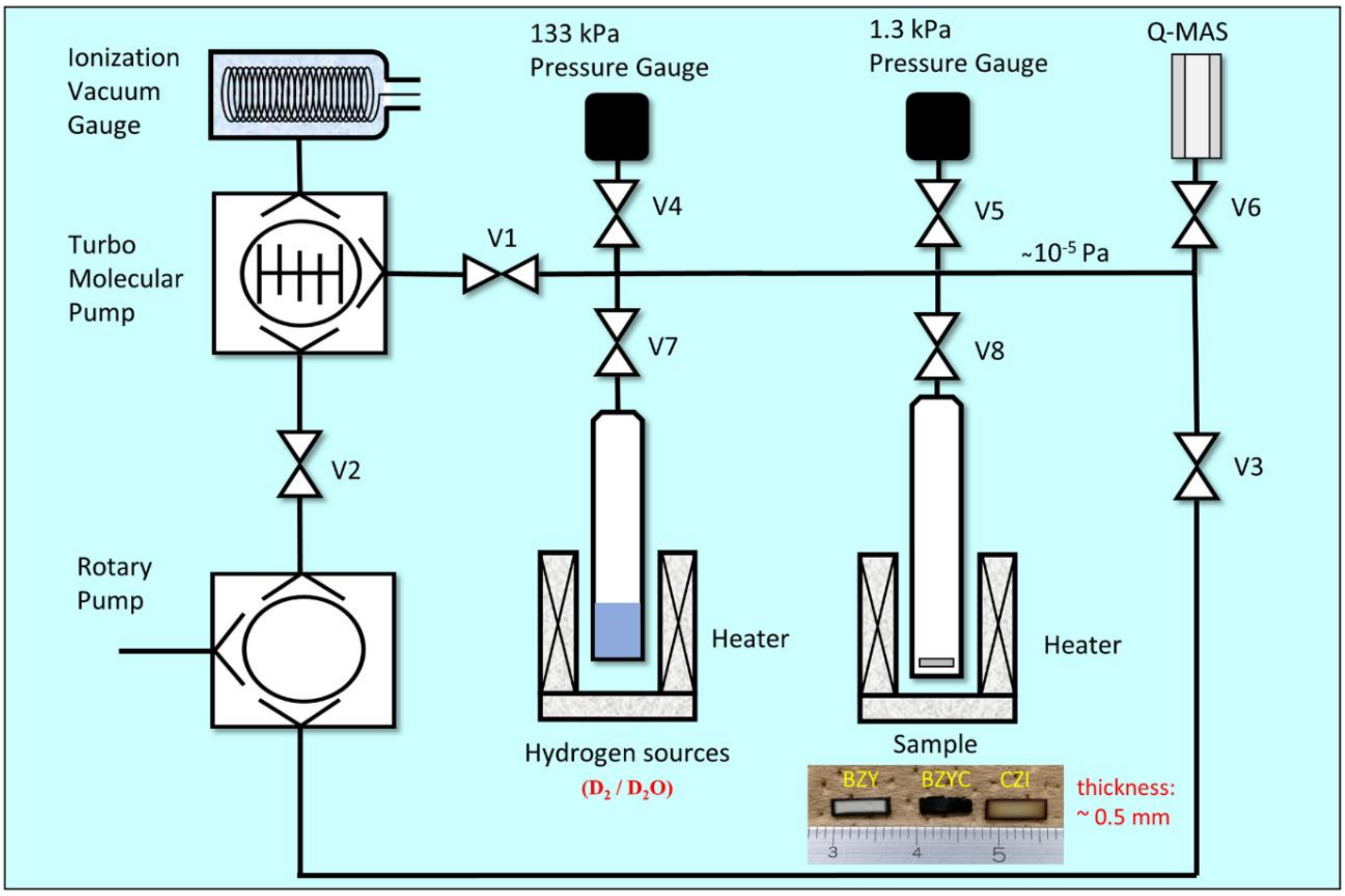

**Figure 1**. Schematic illustration of $D_2$-, or $D_2O$-exposure apparatus, integrated with Q-MAS-baed TDS system.

**Table 1.** List of the $D_2$-exposure conditions utilized in the Set 1 experiment.

| $D_2$-exposure temp. (K) | $D_2$-exposure time (h) | $D_2$-exposure pressure (Pa) | Specimen | Thickness (mm) |
|---|---|---|---|---|
| 673 | 4 | ~ 1333 | BZY | ~ 0.5 |
| | | ~ 1333 | BZYC | ~ 0.5 |
| | | ~ 1333 | CZI | ~ 0.5 |
| 873 | 1 | ~ 1333 | BZY | ~ 0.5 |
| | | ~ 1333 | BZYC | ~ 0.5 |
| | | ~ 1333 | CZI | ~ 0.5 |
| 1073 | 1 | ~ 1333 | BZY | ~ 0.5 |
| | | ~ 1333 | BZYC | ~ 0.5 |
| | | ~ 1333 | CZI | ~ 0.5 |

**Table 2**. List of the $D_2O$-exposure conditions utilized in the Set 2 experiment.

| $D_2O$-exposure temp. (K) | $D_2O$-exposure time (h) | $D_2O$-exposure pressure (Pa)* | Specimen | Thickness (mm) |
|---|---|---|---|---|
| 673 | 4 | ~ 3333 | BZY | ~ 0.5 |
| | | ~ 3333 | BZYC | ~ 0.5 |
| | | ~ 3333 | CZI | ~ 0.5 |
| 873 | 1 | ~ 3333 | BZY | ~ 0.5 |
| | | ~ 3333 | BZYC | ~ 0.5 |
| | | ~ 3333 | CZI | ~ 0.5 |
| 1073 | 1 | ~ 3333 | BZY | ~ 0.5 |
| | | ~ 3333 | BZYC | ~ 0.5 |
| | | ~ 3333 | CZI | ~ 0.5 |

*Saturated water vapor pressure at room temperature.

## 2.3 *TDS experiment*

**Figure 1** shows a schematic view of the $D_2$-gas, or $D_2O$-vapor exposure apparatus, integrated with a Q-MAS-based TDS system. The $D_2$- or $D_2O$-exposed samples were put in the TDS apparatus individually. The temperature of the sample quartz glass tube was then increased at a rate of 0.5 K/s by applying a vacuum pressure of about $10^{-5}$ Pa. A quadrupole mass spectrometer (Q-MAS; TSPTT200, INFICON) was used to measure the gas emitted by the samples [37]. Burnout under the same circumstances without using a sample was done using an electric furnace before performing TDS measurements on a sample exposed to $D_2$- or $D_2O$ to release and eliminate chemicals collected in the instrument. The detailed TDS measurements procedure is discussed in our previous reported papers [15,38].

## 2.4 *Characteristics and principle of Q-MAS*

Mass spectrometry is one of the most popular instrumental analyzers that ionizes and separates chemical substances, such as atoms and molecules, and measures ionized ions by their mass. The major parts of a mass spectrometer are ionization (ion source), mass analysis (analyzer), detection (detector), vacuum exhaust (vacuum pump), data processing (data system), and device control. There are various types of analyzers, one of which is a quadrupole type. A mass spectrometer with a quadrupole analyzer is known as the Q-MAS. Some advantages of Q-MAS are low vacuum (usually ~$10^{-5}$ Torr), miniaturization, high-speed scanning (latest is 10000 *u*/s or more, *u* = unified atomic mass unit), low cost, ease of maintenance and operation, and robustness [39,40].

A Q-MAS consists of four parallel rod-shaped electrodes (**Figure 2**), where a superposition of DC voltage and high-frequency AC voltage is applied to the opposite electrodes having the same polarity to form a quadrupole electric field [39,40]. Here, $U$ is the DC voltage, $\pm (U + \cos \omega t)$ is the AC voltage, $V$ is the maximum value of the AC voltage, $\omega = 2\pi v$, and $v$ is the high frequency. In mass spectrometry, gas molecules are ionized under a high vacuum, and the generated ions are separated and analyzed in the order of the magnitude of the $m/z$ ratio, where $m$ = mass and $z$ = charge. When ions enter the quadrupole electrode, they are affected by the high-frequency electric field and travel in the $z$-direction while oscillating in the $x$ or $y$ direction (**Figure 2**) [39,40]. However, only ions with a specific $m/z$ value produce stable oscillatory motion, pass through the quadrupole, and reach the detector. On the other hand, ions with other $m/z$ values have large amplitudes and diverge and collide with the electrode. When the frequency $f$ is constant and $U/V$ has a specific value, only ions satisfying $m/z = 0.14\ V/f^2r^2$. Therefore, if V is continuously changed while $U/V$ is kept constant, ions corresponding to each mass will be separated and detected. The passing ion stream is usually detected using an electron-amplifier tube. In the electron-amplifier tube, a small number of incident ions hit the metal surface and emit many secondary electrons. By repeating this process, secondary electrons are multiplied, and a small number of ions can be detected with high sensitivity [39,40].

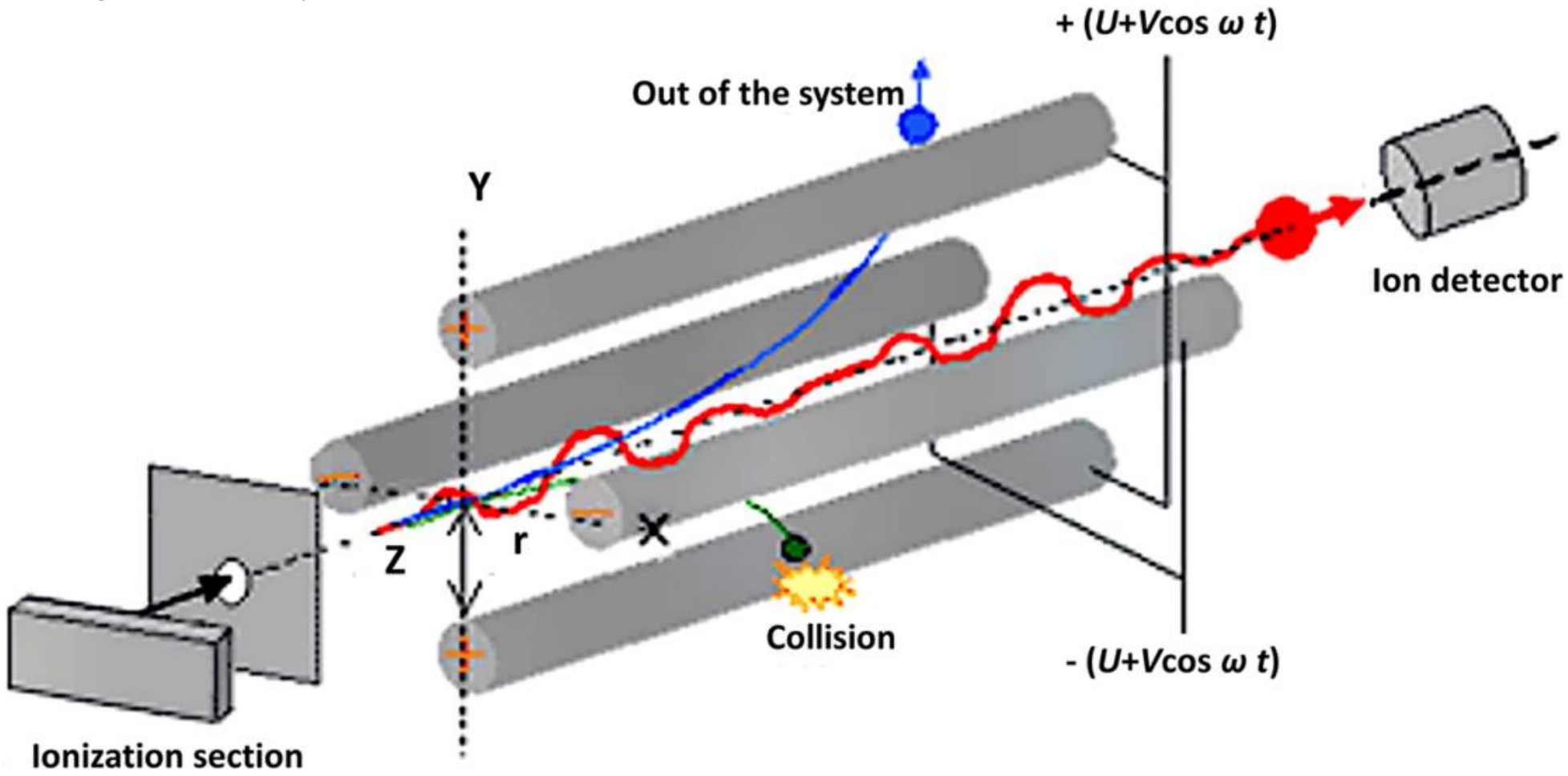

**Figure 2**. Conceptual diagram of ion separation and detection in a quadrupole analyzer [39,40].

## 3 Results and Discussion

### *3.1 Structural and morphological analysis of sintered pellets*

**Figure 3** shows the X-ray diffraction patterns of BZY, BZYC, and CZY powder and sintered bodies. **Table 3** shows the powder and sintered body lattice constants calculated by refining the diffraction pattern for all samples. Based on the values of the lattice constants of each axis and corresponding angles obtained by refining each diffraction pattern, it was confirmed that both the powder and sintered bodies of BZY and BZYC were a single phase of cubic crystals; however, CZI has an orthorhombic structure for both powder and sintered body. BZY and BZYC powders had lattice constants of 4.1967 Å and 4.1939 Å, respectively, whereas BZY and BZYC sintered bodies had lattice values of 4.2128 Å and 4.1986 Å, respectively, which is close to the literature data [41–43].

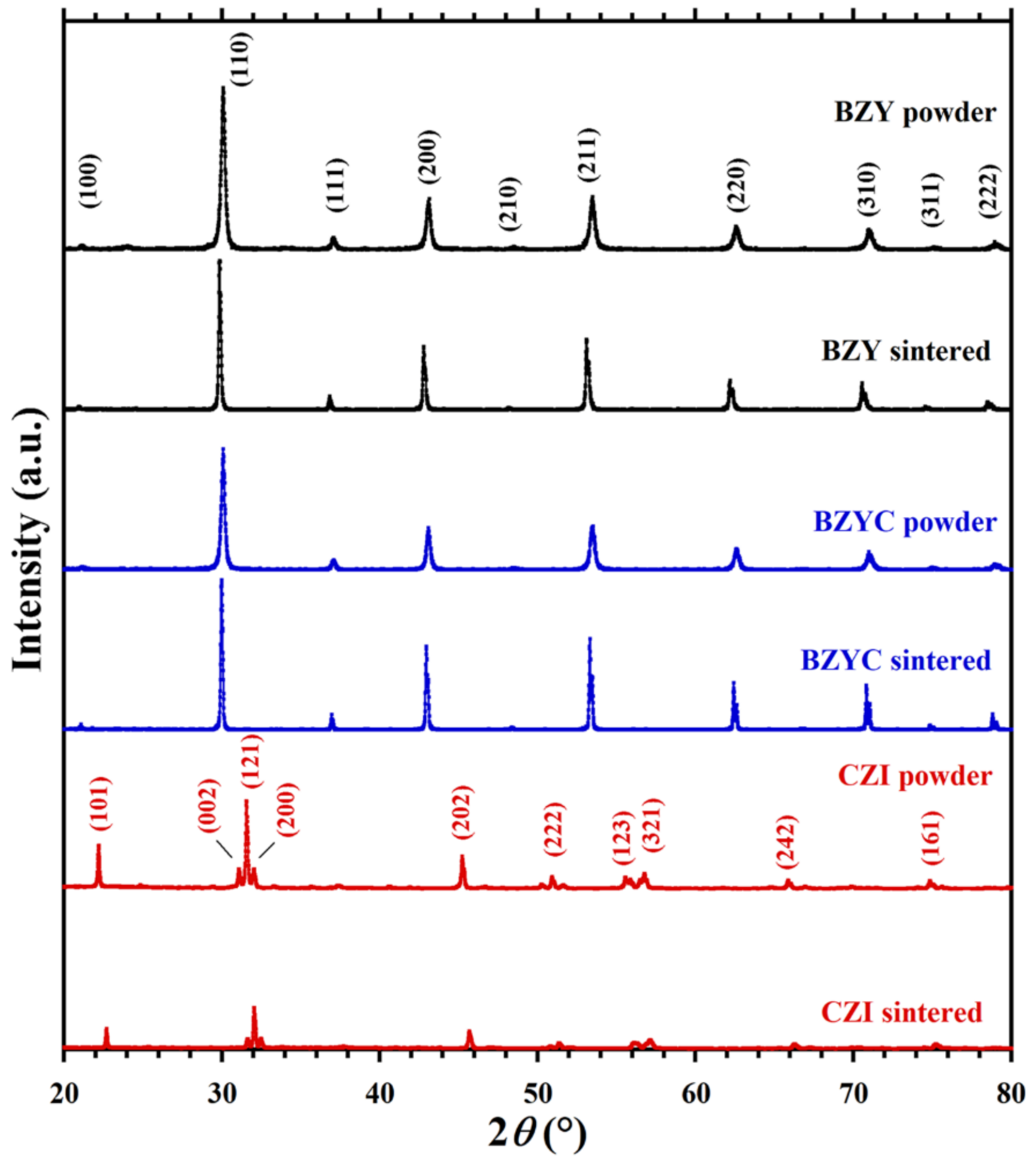

**Figure 3**. XRD patterns for BZY, BZYC, and CZI powder and sintered samples.

**Table 3.** Some parameters of powders and sintered samples.

| Sample | BZY | BZYC | CZI |
|---|---|---|---|
| Powder structure | Cubic | Cubic | Orthorhombic |
| Sintered body structure | Cubic | Cubic | Orthorhombic |
| Powder lattice parameters (Å) | a=b=c=4.1937 | a=b=c=4.1939 | a=5.5916, b=8.0177, c=5.758 |
| Sintered body lattice parameters (Å) | a=b=c=4.2128 | a=b=c=4.1986 | a=6723, b=7.8826, c=5.5716 |
| Avg. particle size of powder (µm) | agglomerated | ~50 | ~100 |
| Avg. grain size of sintered body (µm) | ~0.7 | ~0.85 | ~1.5 |
| Theoretical density (x10$^3$ Kg/m$^3$) | 6.10 | 6.16 | 4.68 |
| Sintered body density (x10$^3$ Kg/m$^3$) | 5.98 | 6.14 | 4.64 |
| % of theoretical density (%TD) | 98.0 | 99.7 | 99.5 |
| Sintered sample open porosity (v%) | 0.1 | 0.1 | 1.3 |
| Sintered sample closed porosity (v%) | 2.1 | 2.5 | 0.5 |

The results of density measurements using the Archimedes method for the BZY, BZYC, and CZY samples are also shown in **Table 3**. BZY, BZYC, and CZY have relative densities of 98%, 99.7%, and 99.5%, respectively. Following the density measurement, it was established that all samples had a very high relative density, and BZYC had a higher density than BZY.

In **Figure 4**, SEM images of BZY, BZYC, and CZI powder and sintered samples are displayed. In addition, after polishing the sample, EDX was used to determine the constituent elements (**Figure 5**). From SEM images (**Figure 4**) it is observed that the BZY powders' particles are agglomerated, and the size of the primary grains is not clearly visible. However, they appear to be 0.1–1.0 μm in size, which is consistent with the literature [44–47]. There were distinct and dense grain boundaries seen in the SEM images of all samples. Cleared grain boundaries and rounded tiny particles are also visible. As can be seen from the fractured surface, it appears that the surface grains were stick together. This means that all chemical components were consistently distributed across its surface, according to EDX mapping (**Figure 5**).

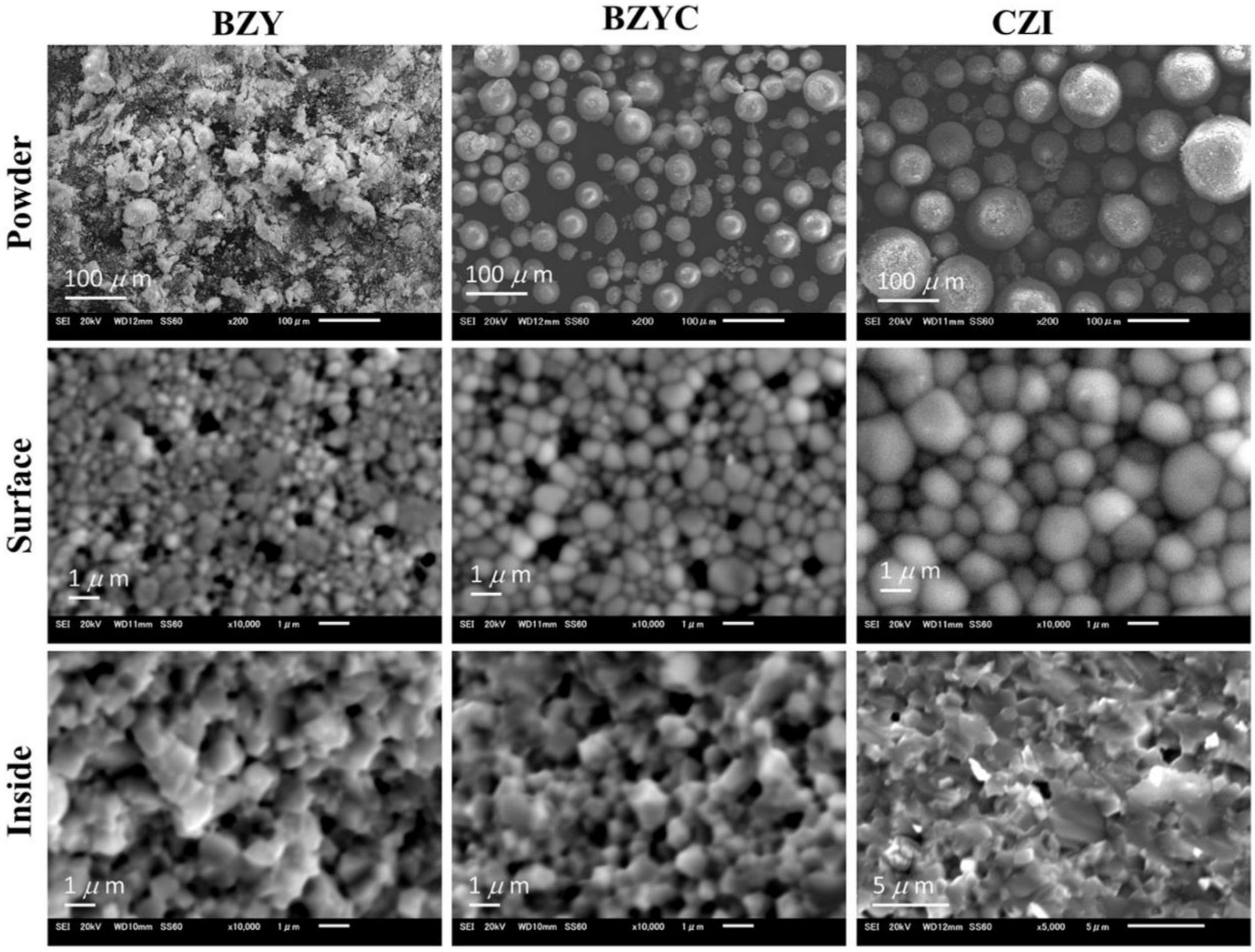

**Figure 4**. BZY, BZYC, and CZI samples' powder, sintered outer surface (OS), and sintered inner (fractured surface) SEM micrographs.

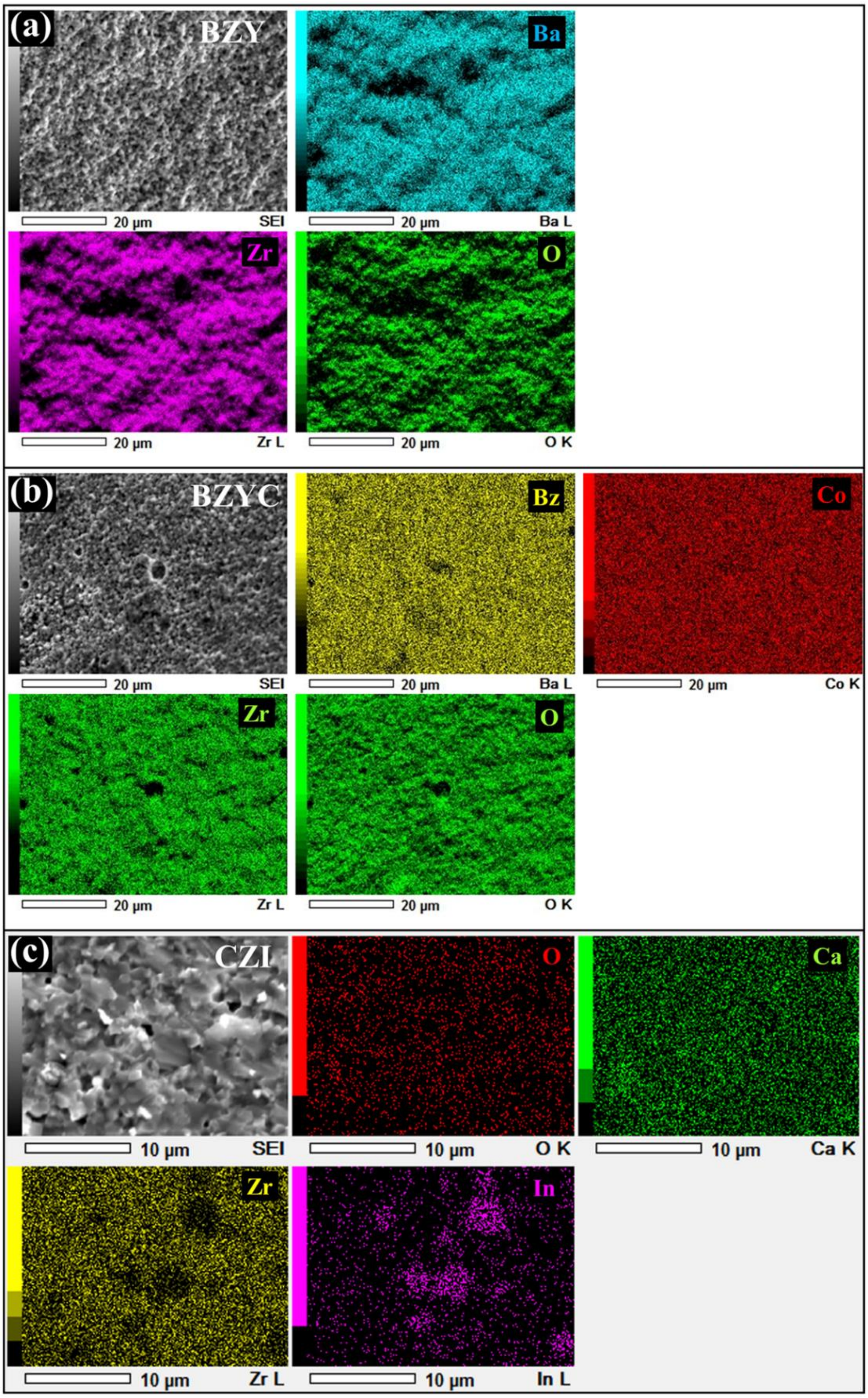

**Figure 5**. Sintered BZY, BZYC, and CZI samples' fractured surface EDX mapping.

### *3.2 Conversion of measured Q-MAS current value to released atoms*

The quantity of hydrogen emitted from $D_2$- or $D_2O$-exposed samples could be calculated using the Q-MAS

measured electric current value. To convert the observed current value obtained from Q-MAS data into the amount of emission, a conventional He leak experiment was done. At ambient temperature, the leak rate of the He standard leak (Model CL-6-He-4FVCR-500DOT-MFV, Vacuum Technology Inc.) $\dot{n}$ was $2.8 \times 10^{-8}$ Pa $m^3 \cdot s^{-1}$. Because the inside of the device was evacuated, it was thought that the gas atmosphere was near to an ideal gas environment. The He standard leak rate $n$ was calculated using the ideal gas equation (Eq. 1) and found to be $1.13 \times 10^{-11}$ mol/s.

$$n = \frac{PV}{RT} \quad (1)$$

If $I_{leak}$ (A) is the current created by the He standard leak, $n_a$ (mol/s) is the release rate of the molecule from the sample, $I_a$ (A) is the current generated by the release of the molecule $a$, and $R_{a/He}$ is the ratio of ionization rates of $a$ and He, then $\dot{n_a}$ may be calculated using Eq (2):

$$\dot{n}_a = \dot{n} \times \frac{I_a}{I_{leak}} \times \frac{1}{R_{a/He}} \quad (2)$$

Since $\mathrm{R_{D_2/He}} = 3, \mathrm{R_{D_2O/He}} = 7.923$ [48], therefore,

$$\dot{n}_{D_2} = \frac{I_{D_2}}{I_{leak}} \times 3.77 \times 10^{-12} [\mathrm{mol/s}] \quad (3)$$

$$\dot{n}_{D_2O} = \frac{I_{D_2O}}{I_{leak}} \times 1.43 \times 10^{-12} [\mathrm{mol/s}] \quad (4)$$

Eqs. (3) and (4) are identical to Eqs. (5) and (6) if the pressure of the He standard leak is $P_{leak}$ (Pa) and the pressure fluctuation during TDS measurement of the molecule $a$ is $P_a$ (Pa):

$$\dot{n}_{D_2} = \frac{I_{D_2}}{I_{leak}} \times \frac{P_{leak}}{P_a} \times 3.77 \times 10^{-12} [\mathrm{mol/s}] \quad (5)$$

$$\dot{n}_{D_2O} = \frac{I_{D_2O}}{I_{leak}} \times \frac{P_{leak}}{P_a} \times 1.43 \times 10^{-12} [\mathrm{mol/s}] \quad (6)$$

In this experiment, Eqs. (5) and (6) were used to calculate the flow rate of molecules of deuterium (HD, HDO, $D_2$, and $D_2O$) from Q-MAS.

### *3.3 Deuterium- and heavy-water-exposure results*

The ion currents obtained from the Q-MAS data for the BZY samples exposed to $D_2$ gas and $D_2O$ vapor are shown in **Figure 6**. **Table 4** provides the hydrogen gas (HD+$D_2$) and water vapor (HDO+$D_2O$) release-peak temperatures for the BZY sample. In BZY samples exposed to $D_2$ at an exposure temperature range of 673–1073 K, hydrogen gas was released at approximately 1010–1090 K, whereas water vapor was released at approximately 1040–1100 K (**Table 4**). In the case of BZY samples exposed to $D_2O$ at an exposure temperature range of 673–1073 K, hydrogen gas was released at approximately 1020–1100 K, whereas water vapor was released at approximately 800–1050 K (**Table 4**).

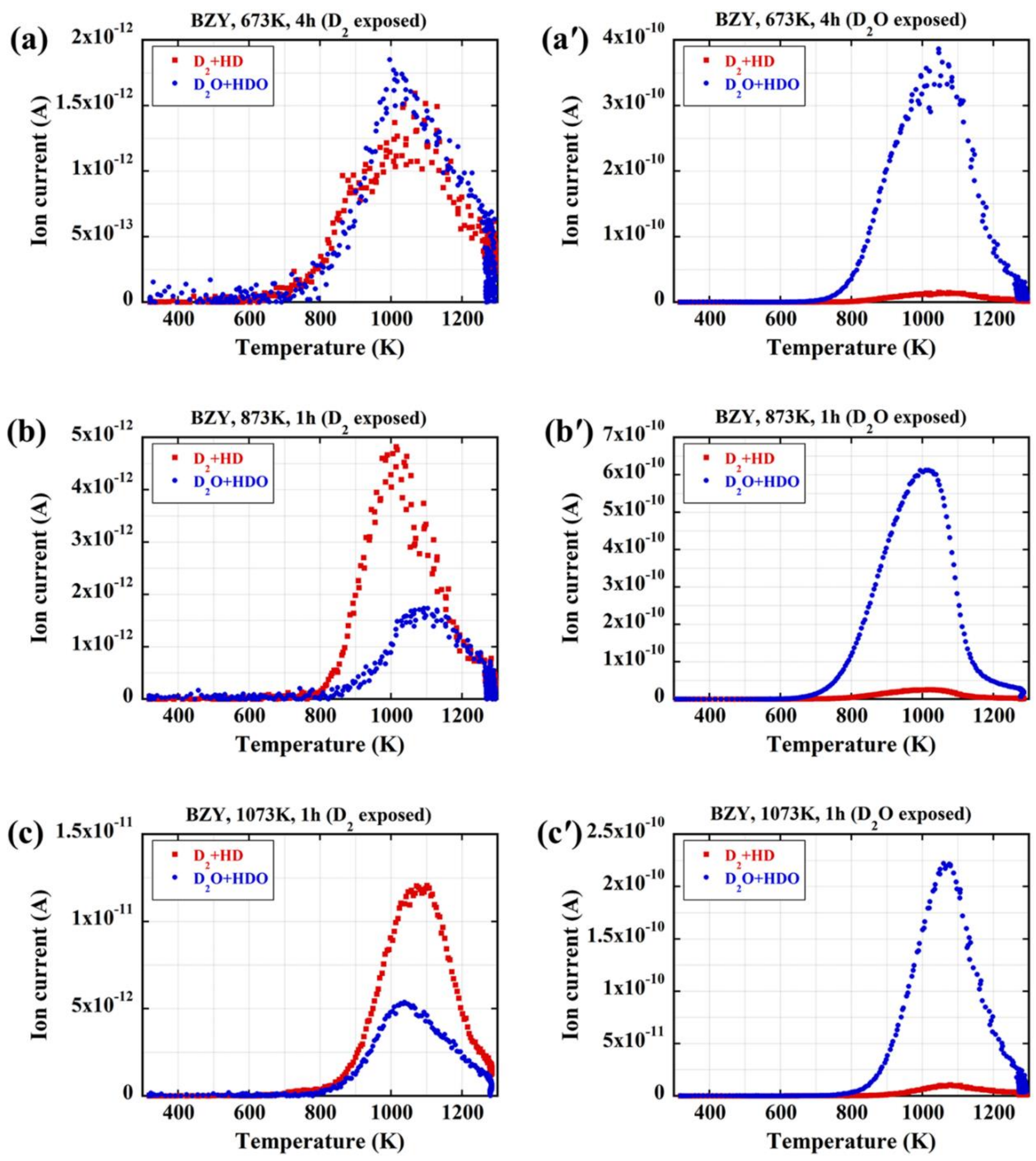

**Figure 6**. Hydrogen gas and water vapor released spectrums for the deuterium- ($D_2$-) and heavy water- ($D_2O$-) exposed BZY sample at 673 K, 873 K, and 1073 K.

**Table 4.** Hydrogen gas and water vapor release-peak temperature for BZY sample.

| Sample | Exposure source | Exposure Temperature | Hydrogen gas ($HD+D_2$) released peak (K) | Water vapor ($HDO+D_2O$) released peak (K) |
|---|---|---|---|---|
| BZY | $D_2$ | 673K, 4h | 1070 | 1000 |
| | | 873K, 1h | 1010 | 1100 |
| | | 1073K, 1h | 1090 | 1040 |
| | $D_2O$ | 673K, 4h | 1050 | 1050 |
| | | 873K, 1h | 1020 | 1020 |
| | | 1073K, 1h | 1100 | 800 |

The ion currents obtained from the Q-MAS data for the $D_2$ gas and $D_2O$ vapor-exposed BZYC samples are shown in **Figure 7**. **Table 5** provides the hydrogen gas (HD+$D_2$) and water vapor (HDO+$D_2O$) release-peak temperatures for the BZYC sample. In the case of BZYC samples exposed to $D_2$ at an exposure temperature range of 673–1073 K, hydrogen gas was released at approximately 820–1050 K, whereas water vapor was released at approximately 980–1110 K (**Table 5**). In the case of BZYC samples exposed to $D_2O$ at an exposure temperature range of 673–1073 K, hydrogen gas was released at approximately 900–1260 K, whereas water vapor was released at approximately 970–1060 K (**Table 5**).

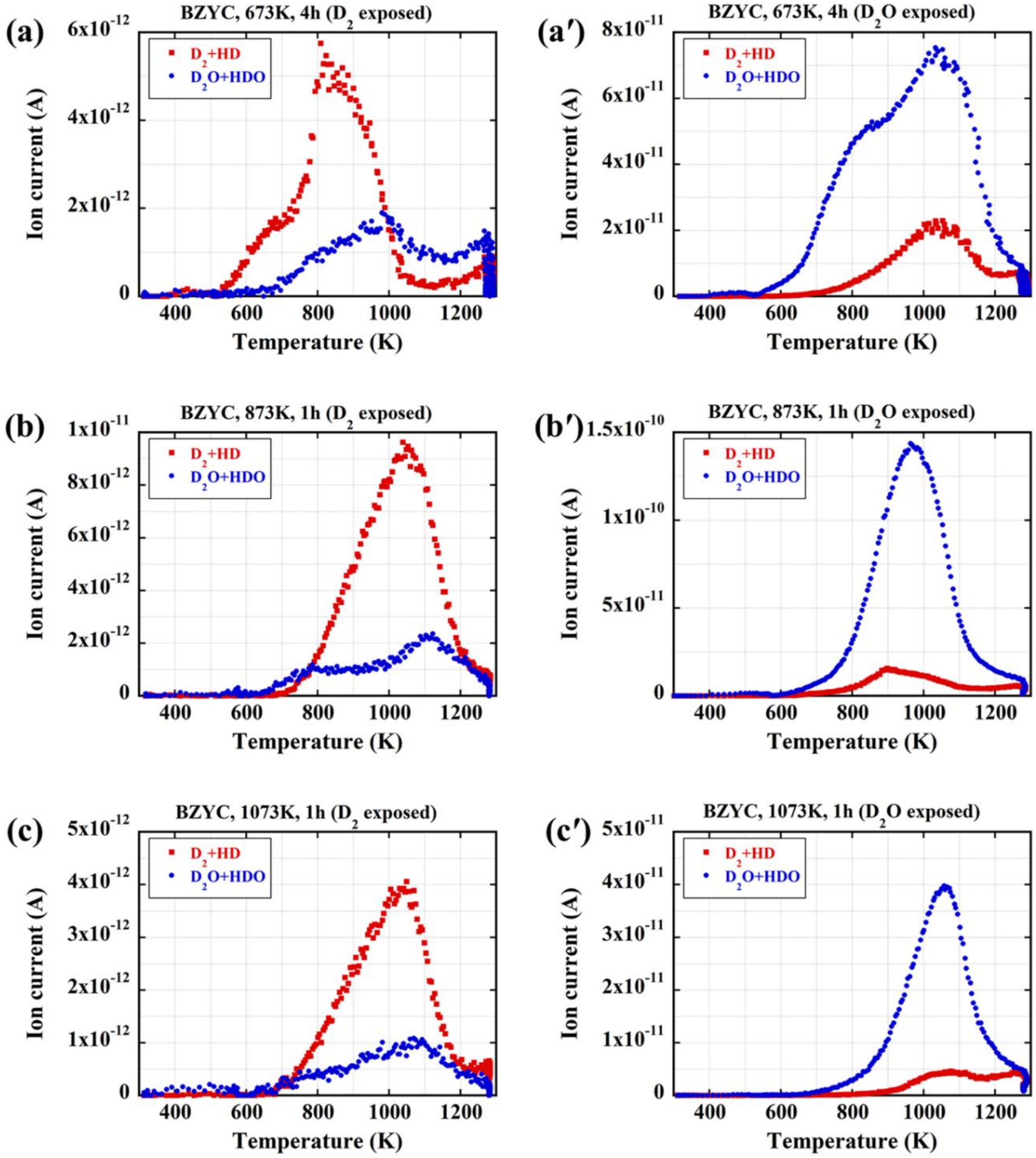

**Figure 7**. Hydrogen gas and water vapor released spectrums for the deuterium- ($D_2$-) and heavy water- ($D_2O$-) exposed BZYC sample at 673 K, 873 K, and 1073 K.

**Table 5.** Hydrogen gas and water vapor release-peak temperature for BZYC sample**.**

| Sample | Exposure source | Exposure Temperature | Hydrogen gas (HD+$D_2$) released peak (K) | Water vapor (HDO+$D_2O$) released peak (K) |
|---|---|---|---|---|
| BZYC | $D_2$ | 673K, 4h | 820 | 980 |
| | | 873K, 1h | 1040 | 1110 |
| | | 1073K, 1h | 1050 | 1070 |
| | $D_2O$ | 673K, 4h | 1040 | 1040 |
| | | 873K, 1h | 900 | 970 |
| | | 1073K, 1h | 1070, 1260 | 1060 |

The ion currents obtained from the Q-MAS data for the $D_2$ gas and $D_2O$ vapor-exposed CZI samples are shown in **Figure 8**. **Table 6** provides the hydrogen gas (HD+$D_2$) and water vapor (HDO+$D_2O$) release-peak temperatures for the CZI sample. In the case of the CZI sample exposed to $D_2$ at 1073 K for 1 h, hydrogen gas was released at approximately 1020 K. Meanwhile, for other exposure conditions (673 K for 4 h or 873 K for 1 h), no hydrogen gas was released with a clear peak (**Table 6**). For the CZI sample exposed to $D_2$ at 673 K for 4 h, water vapor was released at approximately 860 K. Meanwhile, for other exposure conditions (873 K for 1 h or 1073 K for 1 h), no water vapor was released with a clear peak (**Table 6**). In CZI samples exposed to $D_2O$ at 673 K for 4 h, no hydrogen gas or water vapor was released. However, in CZI samples exposed to $D_2O$ at an exposure temperature range of 873–1073 K, hydrogen gas was released at approximately 910–1190 K, whereas water vapor was released at approximately 640–990 K (**Table 6**).

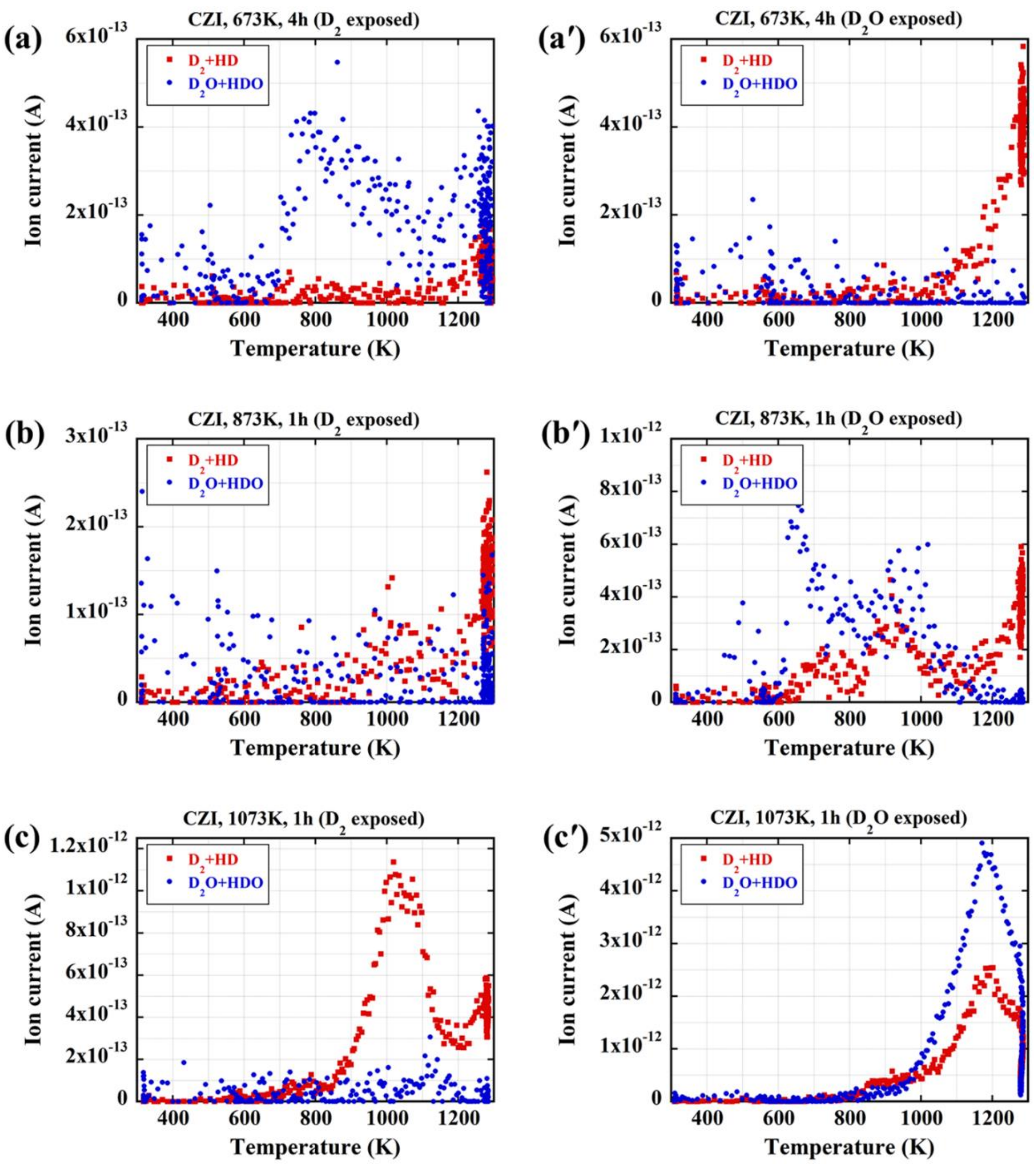

**Figure 8**. Hydrogen gas and water vapor released spectrums for the deuterium- ($D_2$-) and heavy water- ($D_2O$-) exposed CZI sample at 673 K, 873 K, and 1073 K.

**Table 6.** Hydrogen gas and water vapor release-peak temperature for CZI sample**.**

| Sample | Exposure source | Exposure Temperature | Hydrogen gas (HD+$D_2$) released peak (K) | Water vapor (HDO+$D_2O$) released peak (K) |
|---|---|---|---|---|
| CZI | $D_2$ | 673K, 4h | - | 860 |
| | | 873K, 1h | - | - |
| | | 1073K, 1h | 1020 | - |
| | $D_2O$ | 673K, 4h | - | - |
| | | 873K, 1h | 910 | 640, 990 |
| | | 1073K, 1h | 1190 | 1170 |

**Table 7**. Deuterium released during the TDS experiment from $D_2$-exposed samples or Set 1.

| Sample | BZY | | | BZYC | | | CZI | | |
|---|---|---|---|---|---|---|---|---|---|
| | 673K, 4h | 873K, 1h | 1073K, 1h | 673K, 4h | 873K, 1h | 1073K, 1h | 673K, 4h | 873K, 1h | 1073K, 1h |
| Deuterium from $D_2$ and HD (H/M) | 2.1 x $10^{-5}$ | 3.8 x $10^{-5}$ | 1.2 x $10^{-4}$ | 0.7 x $10^{-4}$ | 1.1 x $10^{-4}$ | 5.6 x $10^{-5}$ | 2.0 x $10^{-6}$ | 2.8 x $10^{-6}$ | 1.2 x $10^{-5}$ |
| Deuterium from $D_2O$ and HDO (H/M) | 2.0 x $10^{-5}$ | 1.8 x $10^{-5}$ | 0.4 x $10^{-4}$ | 0.4 x $10^{-4}$ | 0.3 x $10^{-4}$ | 1.9 x $10^{-5}$ | 6.3 x $10^{-6}$ | 1.3 x $10^{-6}$ | 0.1 x $10^{-5}$ |
| Total deuterium (H/M) | 4.1 x $10^{-5}$ | 5.6 x $10^{-5}$ | 1.6 x $10^{-4}$ | 1.1 x $10^{-4}$ | 1.4 x $10^{-4}$ | 7.5 x $10^{-5}$ | 8.3 x $10^{-6}$ | 4.1 x $10^{-6}$ | 1.3 x $10^{-5}$ |
| Fraction of hydrogen gas (%) | 51 | 68 | 75 | 64 | 79 | 75 | 24 | 68 | 92 |
| Fraction of water vapor (%) | 49 | 32 | 25 | 36 | 21 | 25 | 76 | 32 | 8 |

**Table 8**. Deuterium released during the TDS experiment from $D_2O$-exposed samples or Set 2.

| Sample | BZY | | | BZYC | | | CZI | | |
|---|---|---|---|---|---|---|---|---|---|
| | 673K, 4h | 873K, 1h | 1073K, 1h | 673K, 4h | 873K, 1h | 1073K, 1h | 673K, 4h | 873K, 1h | 1073K, 1h |
| Deuterium from $D_2$ and HD (H/M) | 0.2 x $10^{-3}$ | 0.2 x $10^{-3}$ | 0.1 x $10^{-3}$ | 0.4 x $10^{-4}$ | 0.3 x $10^{-3}$ | 1.1 x $10^{-4}$ | 7.2 x $10^{-6}$ | 0.8 x $10^{-5}$ | 2.2 x $10^{-5}$ |
| Deuterium from $D_2O$ and HDO (H/M) | 3.6 x $10^{-3}$ | 5.2 x $10^{-3}$ | 1.7 x $10^{-3}$ | 1.3 x $10^{-4}$ | 1.4 x $10^{-3}$ | 4.3 x $10^{-4}$ | 0.6 x $10^{-6}$ | 0.4 x $10^{-5}$ | 3.1 x $10^{-5}$ |
| Total deuterium (H/M) | 3.8 x $10^{-3}$ | 5.4 x $10^{-3}$ | 1.8 x $10^{-3}$ | 1.7 x $10^{-4}$ | 1.7 x $10^{-3}$ | 5.4 x $10^{-4}$ | 7.8 x $10^{-6}$ | 1.2 x $10^{-5}$ | 5.3 x $10^{-5}$ |
| Fraction of hydrogen gas (%) | 5 | 4 | 5 | 24 | 18 | 20 | 92 | 67 | 42 |
| Fraction of water vapor (%) | 95 | 96 | 95 | 76 | 82 | 80 | 8 | 33 | 58 |

### 3.4 *Hydrogen dissolution and release for $D_2$ and $D_2O$ exposure*

The deuterium amounts released from all samples (BZY, BZYC, and CZI) for $D_2$ and $D_2O$ exposure are given in **Table 7** and **Table 8**, respectively. Based on **Table 7** and **Table 8**, **Figure 9** was drawn. For $D_2$- and $D_2O$-exposed BZY, BZYC, and CZI samples, **Figure 9** summarizes the atomic ratio between hydrogen and metal atoms. The proportions of hydrogen gas ($D_2$ + HD) and water vapor ($D_2O$ + HDO) are shown in the red and blue parts of the pie chart in **Figure 9**, respectively.

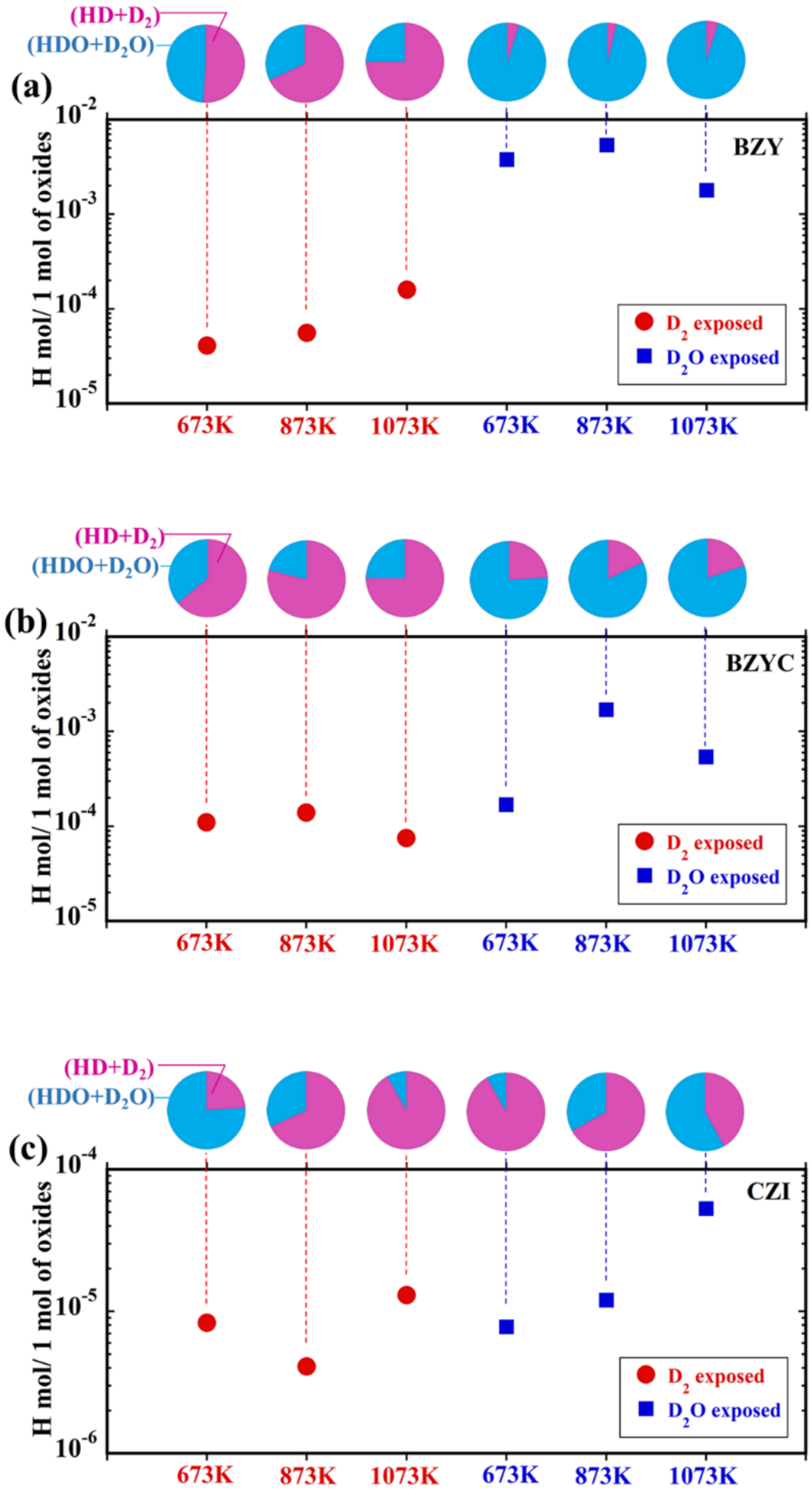

**Figure 9**. For $D_2$- and $D_2O$-exposed (a) BZY, (b) BZYC, and (c) CZI samples, the atomic ratio between hydrogen and metal atoms was summarized. The proportion of hydrogen gas ($D_2$ + HD) is shown in red, while the proportion of water vapor ($D_2O$ + HDO) is shown in blue.

From **Figure 9(a)**, it is clear that, for the case of the $D_2O$-exposed BZY sample, the dissolution of hydrogen is one order of magnitude higher than the $D_2$-exposure case in the temperature range of 673–1073 K. For the $D_2O$-exposure cases, most of the dissolved hydrogen was released as water vapor ($D_2O$ + HDO) as compared to hydrogen gas ($D_2$ + HD) release, but opposite results were found for the $D_2$-exposed samples. In the case of the BZYC sample, a similar trend of hydrogen dissolution was observed (**Figure 9(b)**). A higher amount of hydrogen was dissolved in both BZY and BZYC under $D_2O$ exposure at 873 K, but the highest amount was found for BZYC. On the other hand, for both $D_2$- and $D_2O$-exposed CZI samples, an almost similar order of hydrogen dissolution was observed, but this amount was consistently lower than that of BZY and BZYC (**Figure 9(c)**). In addition, in most cases, dissolved hydrogen is released as water vapor ($D_2O$ + HDO) in the CZI samples. These TDS results suggest that CZI is not preferable for electrochemical applications because of low hydrogen dissolution compared to BZY and BZYC. Under $D_2O$ exposure at 873 K, Y, and Co co-doped barium zirconate, that is, BZYC might be preferable for electrochemical applications due to its higher hydrogen dissolution.

**Figure 10(a)-(c)** shows the Arrhenius plot of hydrogen solubility in the temperature range of 673 K to 1073 K for $D_2$- and $D_2O$-exposed BZY, BZYC, and CZI samples, obtained from the TDS data. A similar trend of temperature-dependent hydrogen solubility was obtained for the HT- and DTO-exposure cases (**Figure 10(d)**) [11,12]. Therefore, it could be concluded that the hydrogen solubility data obtained in these studies from $D_2$, $D_2O$, HT, and DTO exposure using TDS are consistent with previously reported tritium exposed TIP data [11,12].

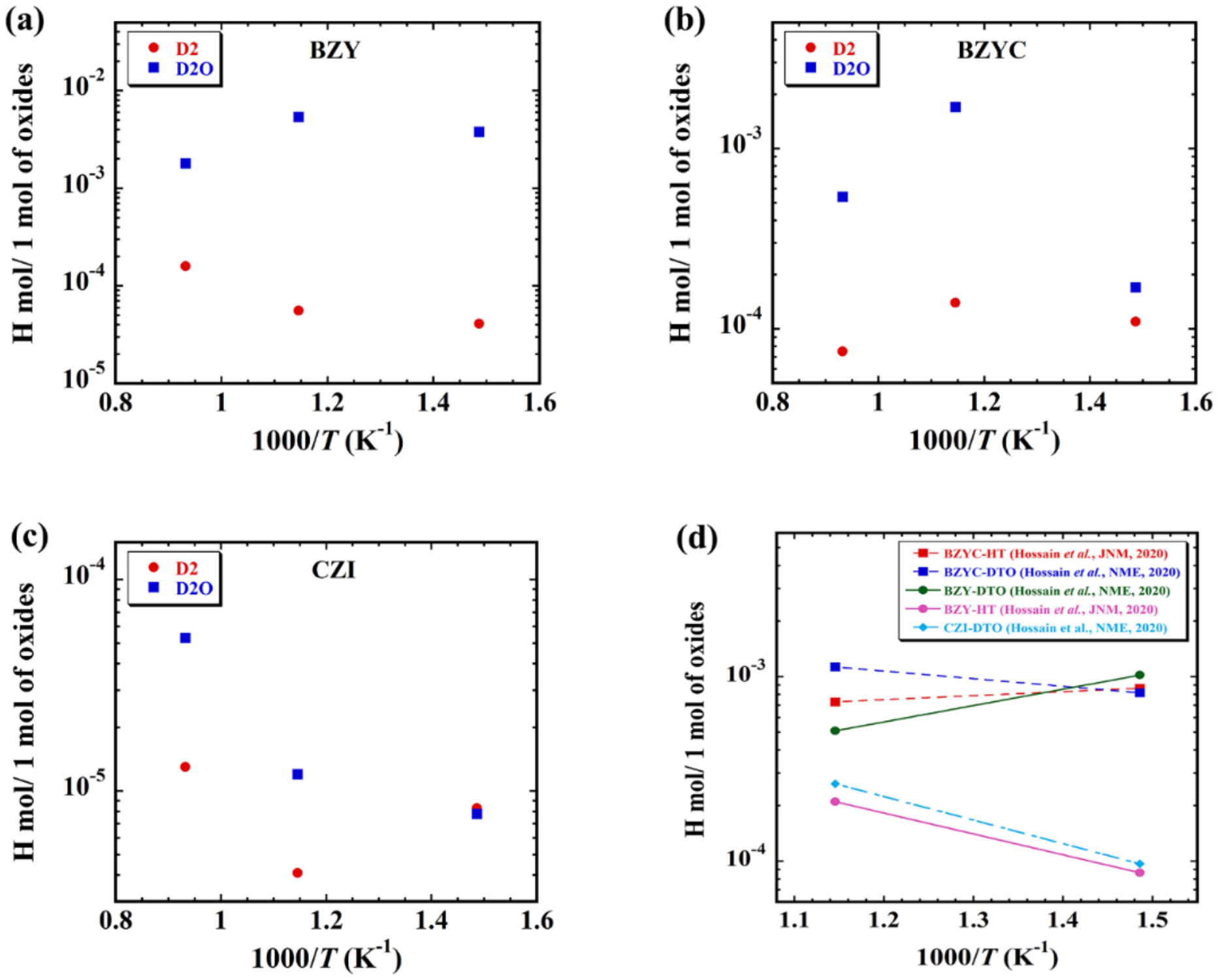

**Figure 10**. Arrhenius plot of hydrogen solubilities in the temperature range of 673 K to 1073 K for $D_2$- and $D_2O$-exposed (a) BZY, (b) BZYC, and (c) CZI samples, obtained from TDS data. (d) Arrhenius plot of hydrogen solubilities in the temperature range of 673 K to 873 K for the BZY, BZYC [12], and CZI [11] samples obtained for the HT- and DTO-exposure from TIP data..

### 3.5 *Comparison between TDS data with TMAP4's simulated data*

The TMAP4 simulated data was compared to the TDS experimental data as well as our previous reported experimental TIP data [11,12]. By specifying the initial and boundary conditions in TMAP4 (a one-dimensional calculation tool for the tritium diffusion process), it is possible to simulate the hydrogen release behavior from the sample (if the sample is sandwiched between two containers or enclosures). The assumptions used in the TMAP4 simulation of the TDS results were: (1) hydrogen is uniformly dissolved in the sample at the beginning (*initial condition*), (2) the pressure in the container is zero, and the surface concentration is zero (*boundary condition*), and (3) hydrogen atoms are the diffusing species, whereas hydrogen molecules are the component species in the container (*species*).

The diffusivity was changed in the TMAP4 simulation to replicate the hydrogen release behavior, which was similar to the experimental results. Eq. (7) can be used to express the diffusivity, $D$ ($m^2/s$):

$$D = D_0 \ exp\left(-\frac{Q}{k_B T}\right) \tag{7}$$

Here $D_0$ = interstitial diffusion ($m^2/s$), $Q$ = activation energy (eV), $k_B$ = Boltzmann constant ($8.617 \times 10^{-5}$ eV $K^{-1}$), and $T$ = sample temperature (K). Eq. (8) expresses the interstitial diffusion $D_0$:

$$D_0 \geqq a_0^2\gamma \simeq 10^{-7} \ m^2/s \ [49] \tag{8}$$

Here, $a_0$ = jump distance (m), and $\gamma$ = hydrogen vibration frequency (/s). In the equation Eq. (7), the ion current spectrum and hydrogen release behavior from the hydrogen-exposed sample were obtained by changing simply the value of the activation energy $Q$.

**Table 9.** $D_0$ and $Q$ values consider for TMAP4 simulation.

| Considering H-atmosphere | Specimen | $D_0$ ($m^2/s$) | Activation energies, $Q$ (eV) |
|---|---|---|---|
| Gaseous | BZY | $1.4 \times 10^{-6}$ | 0.15, 0.20, 0.30 |
| | BZYC | $1.1 \times 10^{-9}$ | 0.15, 0.20, 0.30 |
| | CZI | $2.0 \times 10^{-9}$ | 0.20, 0.30, 0.40 |
| Wet | BZY | $2.0 \times 10^{-9}$ | 0.20, 0.30, 0.40 |
| | BZYC | $6.1 \times 10^{-9}$ | 0.20, 0.30, 0.40 |
| | CZI | $7.3 \times 10^{-8}$ | 0.50, 0.60, 0.70 |

In Eq. (7) if we consider $D_0$ as a constant, then by changing the activation energy $Q$ we can get different hydrogen release behavior curves from TMAP4 simulation. **Table 9** shows the values of $D_0$ which we consider for different samples (BZY, BZYC, and CZI) in various atmospheres (H- gaseous or –wet atmospheres). Using the $D_0$ of **Table 9** and for different activation energies (0.15, 0.20, 0.30, 0.40, 0.50, 0.60, and 0.70 eV), we can get different hydrogen release behavior curve as shown in **Figure 11**. Activation energy, $Q$ vs. hydrogen release peak temperature curve obtained from TMAP4 simulated data for various $Q$ values are shown in **Figure 12**. TMAP4 simulation was carried out by choosing suitable activation energy values so that the simulated emission peak temperature values match with the experimental peaks, as shown in **Figure 13**. Now using the $D_0$ and $Q$ values (**Table 10**), which is considered for TMAP4 simulation of **Figure 13**, from Eq. (7) we can calculate the *D* values which are listed in **Table 10**. By comparing the literature's $D_0$ and *Q* values with current TMAP4 simulated values we found that in most of the cases the simulated *Q* and *D* values are higher than the literate ones. That's because in the sample the diffusion of hydrogen molecules is strongly affected by the diffusion of oxygen, while in this current TMAP4 simulation, only hydrogen molecules are considered [38].

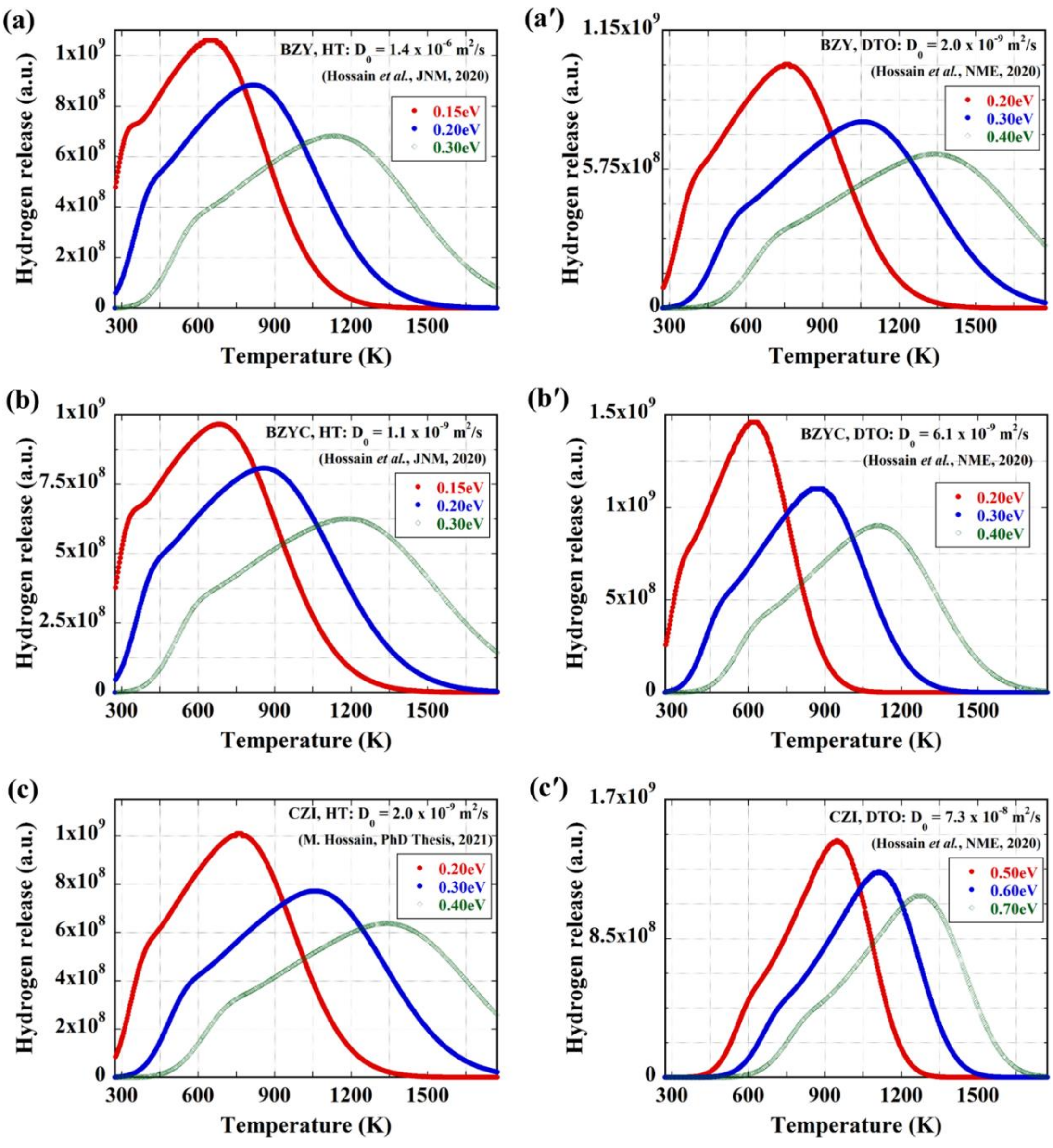

**Figure 11**. TMAP4 simulated hydrogen release behavior curve for various activation energies by considering $D_0$ value obtained from HT exposure (or H-gaseous atmosphere) experiments for (a) BZY [12], (b) BZYC [12], and (c) CZI [50]. TMAP4 simulated hydrogen release behavior curve for various activation energies by considering $D_0$ value obtained from DTO exposure (or H-wet atmosphere) experiments [11] for (a′) BZY, (b′) BZYC, and (c′) CZI.

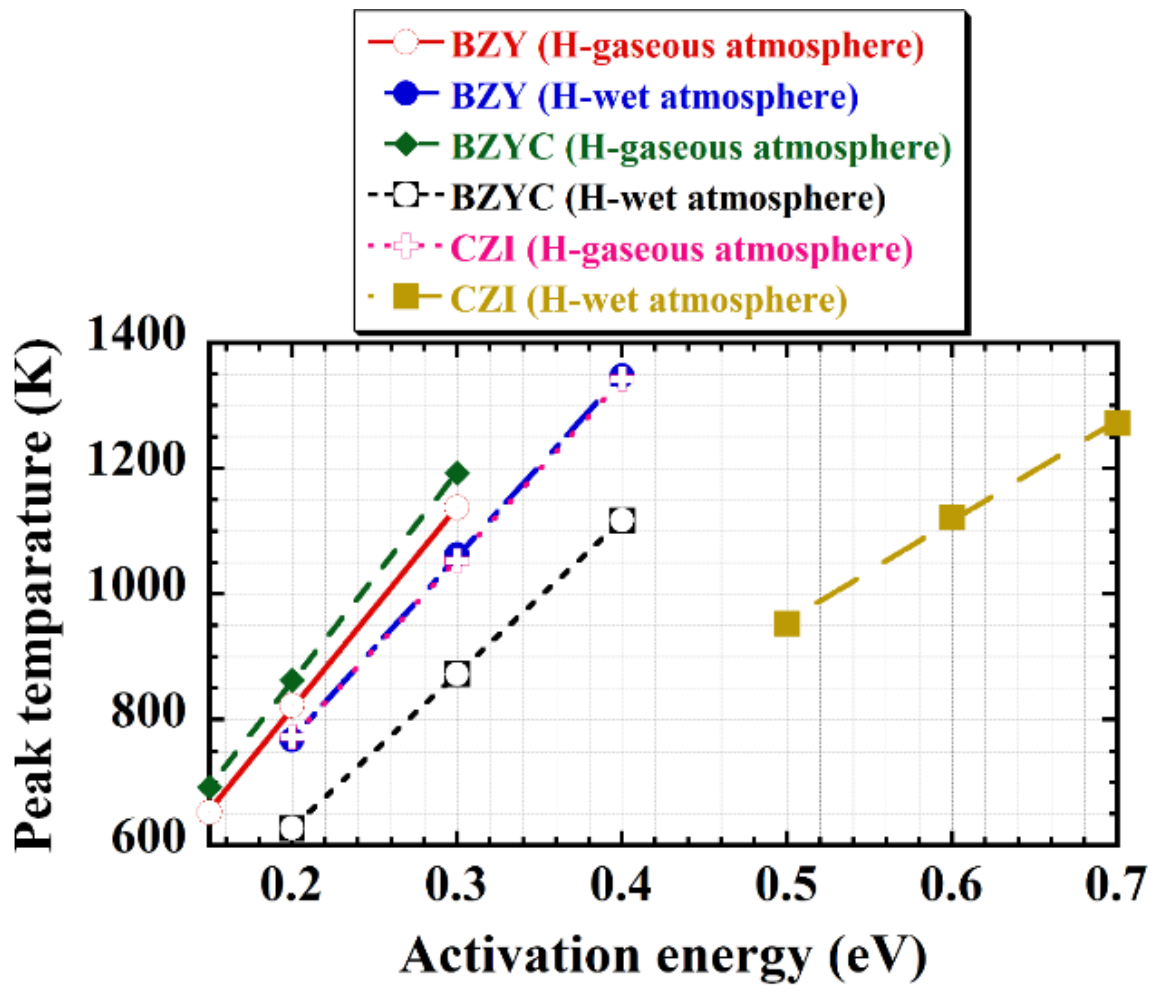

**Figure 12**. Activation energy, $Q$ vs. hydrogen release peak temperature curve obtained from TMAP4 simulation data for various $Q$ values.

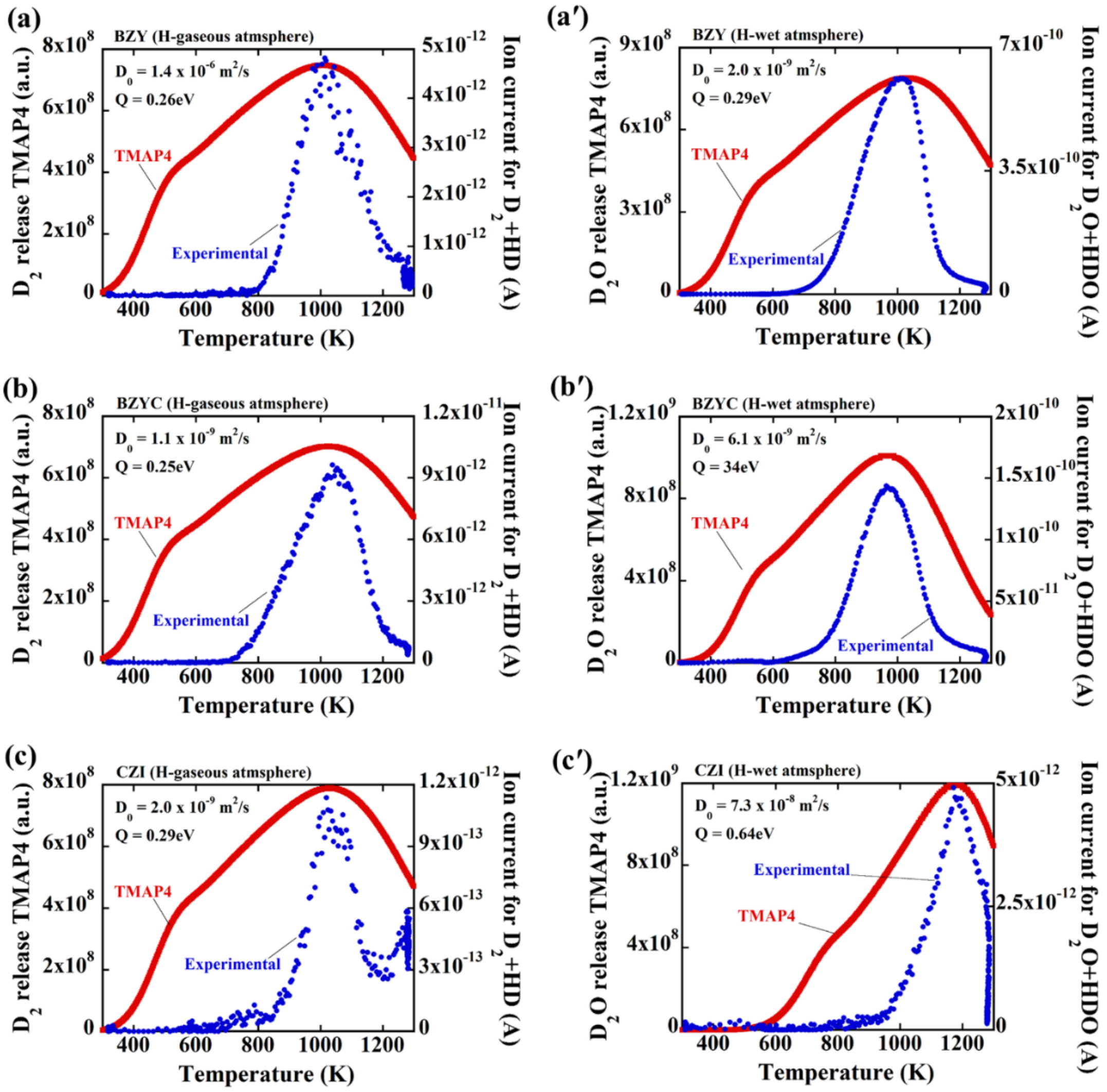

**Figure 13**. TMAP4 simulated hydrogen release behavior curve matching with current TDS experiments' $D_2$-exposed D+HD released peak by considering $D_0$ value obtained from HT exposure (or H-gaseous atmosphere) experiments for (a) BZY (set

Q = 26eV) [12], (b) BZYC (set Q = 25eV) [12], and (c) CZI (set Q = 29eV) [50]. TMAP4 simulated hydrogen release behavior curve matching with current TDS experiments' $D_2O$-exposed $D_2O$+HDO released peak by considering $D_0$ value obtained from DTO exposure (or H-wet atmosphere) experiments [11] for (a′) BZY (set Q = 29eV), (b′) BZYC (set Q = 34eV), and (c′) CZI (set Q = 64eV).

**Table 10.** TMAP4 simulation results compare with literature data [11,12] for an exposure temperature of 873 K.

| Considering H-atmosphere | Specimen | $D_0$ (m²/s) | Set activation energy, *Q* (eV) | Literature activation energy, *Q* (eV) [11,12] | Calculated *D* from TMAP4 (m²/s) | Literature *D* from TIP (m²/s) [11,12] |
|---|---|---|---|---|---|---|
| Gaseous | BZY | $1.4 \times 10^{-6}$ | 0.26 | 0.74 | $4.4 \times 10^{-8}$ | $(7.0 \pm 1.3) \times 10^{-11}$ |
| | BZYC | $1.1 \times 10^{-9}$ | 0.25 | 0.18 | $4.0 \times 10^{-11}$ | $(9.8 \pm 1.3) \times 10^{-11}$ |
| | *CZI | $2.0 \times 10^{-9}$ | 0.29 | 0.32 | $1.4 \times 10^{-10}$ | $(2.8 \pm 0.0) \times 10^{-11}$ |
| Wet | BZY | $2.0 \times 10^{-9}$ | 0.29 | 0.23 | $4.2 \times 10^{-11}$ | $(9.4 \pm 0.1) \times 10^{-11}$ |
| | BZYC | $6.1 \times 10^{-9}$ | 0.34 | 0.31 | $6.6 \times 10^{-11}$ | $(6.25 \pm 0.75) \times 10^{-11}$ |
| | *CZI | $7.3 \times 10^{-8}$ | 0.64 | 0.54 | $2.1 \times 10^{-10}$ | $(5.5 \pm 0.5) \times 10^{-11}$ |

*At 1273 K.

## 4 Summary

In this experiment, BZY, BZYC, and CZI samples were exposed to $D_2$ or $D_2O$ vapor for the TDS experiment. In the $D_2O$-exposed BZY sample, the dissolution of hydrogen was one order of magnitude higher than that of the $D_2$-exposure case in the temperature range of 673–1073 K. For the $D_2O$-exposure cases, most of the dissolved hydrogen was released as water vapor ($D_2O$ + HDO) as compared to hydrogen gas ($D_2$ + HD) release, but opposite results were found for the $D_2$-exposed samples. In the case of the BZYC sample, a similar trend of hydrogen dissolution was observed. A higher amount of hydrogen was dissolved in both BZY and BZYC under $D_2O$ exposure at 873 K, but the highest amount was found for BZYC.

For both $D_2$- and $D_2O$-exposed CZI samples, a similar order of hydrogen dissolution (~$10^{-5}$ H mol/1 mol of oxides) was observed at lower exposure temperature (673 K). However, at higher exposure temperature (≥873 K) $D_2O$-exposed sample shows almost one order higher solubility than the $D_2$-exposure case. The amount of dissolved hydrogen in CZI was consistently lower than that in BZY and BZYC. In addition, in most cases, the dissolved hydrogen is released as water vapor ($D_2O$ + HDO) from the CZI samples. These TDS results suggest that, because of low hydrogen dissolution, CZI is not preferable for electrochemical applications as compared to BZY and BZYC. Under $D_2O$ exposure at 873 K, Y, and Co co-doped barium zirconate, that is, BZYC might be preferable for electrochemical applications owing to its higher hydrogen dissolution.

In most of the cases, the hydrogen diffusion activation energies and diffusivities calculated by TMAP4 were higher than the experimental values, because oxygen is not considered a diffusion species in TMAP4. From the Arrhenius plot of hydrogen solubility obtained from the TDS study, in the temperature range of 673 K to 1073 K for $D_2$- and $D_2O$-exposed BZY, BZYC, and CZI samples, a similar trend of temperature-dependent hydrogen solubility was obtained, similar to the HT- and DTO-exposure cases. Therefore, it could be said that the hydrogen solubility data obtained in these studies from $D_2$, $D_2O$, HT, and DTO exposure are consistent.

## Acknowledgments

This work was financially supported by Kyushu University, Japan. We sincerely appreciate the support and aid received from the scholarship (for M.K. Hossain) given by the Ministry of Education, Culture, Sports, Science and Technology, Japan (MEXT).

## Declaration of interests

The authors declare that they have no known competing financial interests or personal relationships that could have appeared to influence the work reported in this paper.

## Data availability

The raw/processed data required to reproduce these findings cannot be shared at this time as the data also forms part of an ongoing study.